\documentclass[runningheads]{llncs}
\usepackage[T1]{fontenc}
\usepackage{xurl}
\usepackage{multirow}
\usepackage{algorithmic}
\usepackage{graphicx}
\usepackage{textcomp}
\usepackage{comment}
\usepackage{xcolor}
\usepackage{textcomp}
\usepackage{xcolor}
\usepackage{amssymb}
\usepackage{verbatim}
\usepackage{multirow}
\usepackage{array}
\usepackage{longtable}
\usepackage{booktabs}
\usepackage{tabularray}
\usepackage{hyperref}
\usepackage{caption}
\usepackage{tabularx}
\usepackage{graphicx}
\usepackage{multicol}
\usepackage{lscape}
\usepackage{pdflscape}
\usepackage{xcolor}
\usepackage{float}
\usepackage{comment}
\usepackage{tcolorbox}
\usepackage{amsmath}
\usepackage{graphicx}
\begin{document}
\title{\textit{SmartGSN}: an online tool to semi-automatically manage assurance cases}
%
%
\author{Oluwafemi Odu\inst{1}  \and
Daniel Méndez Beltrán\inst{1,2} \and
Emiliano Berrones Gutiérrez\inst{1,2} \and
Alvine B. Belle \inst{2}
\and
Gerhard Yu\inst{1}
\and
Melika Sherafat\inst{1}}
\authorrunning{O. Odu et al.}
%
\institute{Lassonde School of Engineering at York University, Toronto, Canada 
\email{\{olufemi2@, alvine.belle@lassonde.,  gerhardy@, meddeta@my.\}yorku.ca}\\
\and
Monterrey Institute of Technology and Higher Education, Mexico, Mexique\\
\email{\{a01369091, a01770621\}@tec.mx}\\
}
\maketitle              
\begin{abstract}
Developing industry-wide standards and ensuring producers of mission-critical systems comply with them is crucial to fostering consumer acceptance. Producers of such systems can rely on assurance cases to demonstrate to regulatory authorities how they have complied with such standards to help prevent system failure, which could result in fatalities and environmental damage. 
In this paper, we introduce \textit{SmartGSN}, an innovative online tool that relies on Large Language Models to (semi-)automate the management of assurance cases complying with GSN —a very popular notation to graphically represent assurance cases. 
The evaluation of \textit{SmartGSN} demonstrates its strong capability to detect assurance case patterns within the assurance cases manually created for five systems spanning several application domains.  \textit{SmartGSN} is accessible online at [{https://smartgsn.vercel.app}], and a demonstration video can be viewed at [{https://youtu.be/qLrTHf-SZbM}].

\keywords{Assurance and certification \and Assurance cases \and Assurance case patterns \and Large language models \and Generative AI \and GPT.}
\end{abstract}

\section{Introduction}
The producers of mission-critical systems designed to perform essential tasks must usually demonstrate that their systems effectively support their intended non-functional properties (e.g., safety, and security) \cite{b5}. This allows certifying these systems comply with specific industrial standards and allows regulators to authorize the deployment of these systems \cite{b5}. System assurance supports that demonstration by relying on assurance cases (ACs). An assurance case is a well-established, structured, reasoned, and auditable set of arguments supported by a body of evidence and designed to demonstrate
that the non-functional requirements of a system have been correctly implemented \cite{b7}. 
Using assurance cases is common across various application domains (e.g., automotive) since it allows preventing system failure. The latter could result in catastrophic outcomes.


Assurance cases can be represented using either textual notations (e.g., structured prose) or graphical notations \cite{b8}. 
The most used graphical notation to represent assurance cases is the Goal Structuring Notation (GSN) \cite{b9}. Assurance cases represented in GSN are depicted in a tree-like structure called the \textit{goal structure} \cite{b9}. The latter provides a diagrammatic representation of assurance cases. Assurance case patterns (ACPs) are reusable templates that facilitate the creation of assurance cases from recurring assurance case structures \cite{b9}.  

System assurance management activities include: the construction 
 of ACs, the instantiation of ACs, their assessment, their formalization, and their formal verification \cite{b5}. 
However, manually creating and maintaining these assurance cases can be tedious, error-prone, and time-consuming especially since assurance cases are usually very large documents that may consist of several hundred pages \cite{b5}, \cite{b6}. Thus, several tools (e.g., MMINT-A \cite{b20}, Trusta \cite{b21}, SPIRIT \cite{b25}, ExplicitCase \cite{b26}, and Advocate \cite{b19}) support the management  of assurance cases. These tools range from simple graphic editors for drawing assurance cases to comprehensive platforms offering advanced features, such as traceability between assurance case elements, support for multiple assurance case notations, and type checking to validate the content and structure of assurance cases. 
Maksimov et al. \cite{b5} surveyed some of these tools with an emphasis on assurance case assessment.


To the best of our knowledge, there is currently no online tool that leverages Generative Artificial Intelligence (AI) through Large Language Models (LLMs) for the semi-automatic management of GSN-compliant assurance cases. The lack of such tools hinders the automation of the assurance case development process, forcing users to depend on manual, tedious, and time-consuming methods. Without natural language processing capabilities, assurance case developers continue to face increased workloads as they must manually extract system artifacts from system models and replace abstract information within patterns to construct system-specific assurance cases. To bridge that gap, we propose \textit{SmartGSN}, a novel online tool relying on LLMs to support the semi-automatic management of GSN-compliant assurance cases. Its envisioned users are researchers and practitioners (e.g., regulatory authorities, corporate safety  analysts, manufacturers).

The contributions of this paper are three-fold: 1) we present an online tool that relies on LLMs to manage GSN-compliant assurance cases; 2) we describe the technological stack we used to implement and deploy that tool; and 3) we report an evaluation of the proposed tool. Results from that evaluation showed that our tool is performant in detecting patterns within assurance cases. 

In the remainder of this paper, we describe our tool and report an empirical validation of one of its core features.

\section{Background and Related Work}

\subsection{Assurance cases and assurance case patterns}
Assurance cases allow supporting system assurance. Assurance cases are structured arguments with a supporting body of evidence that allow demonstrating that the non-functional requirements of a system have been correctly implemented \cite{b4}. This allows for preventing system failure. The latter can have catastrophic consequences such as the death of people, injuries, and environmental damages. There are several types of assurance cases depending on the system requirement they target: safety cases, security cases, etc.

Several notations allow representing assurance cases. These include graphical notations (e.g, GSN \cite{b9}, CAE \cite{b4}, Eliminative Argumentation \cite{b30}) and textual notations (e.g., structured prose \cite{b8}, unstructured prose \cite{b8}). GSN is the most popular graphical notation. We therefore focus on it in this paper. The main elements that allow the representation of an assurance case in GSN are \cite{b9}:

\begin{enumerate}
    \item \textbf{Goal}: is depicted with a rectangle and, represents a claim forming part of the argument.
    \item \textbf{Strategy}: is represented with a parallelogram and, describes the inference that exists between a goal and its supporting goals.
    \item \textbf{Solution}: is rendered as a circle and, refers to an evidence item.
\end{enumerate}

\begin{figure}
      \centering
      \includegraphics[width=1\linewidth]{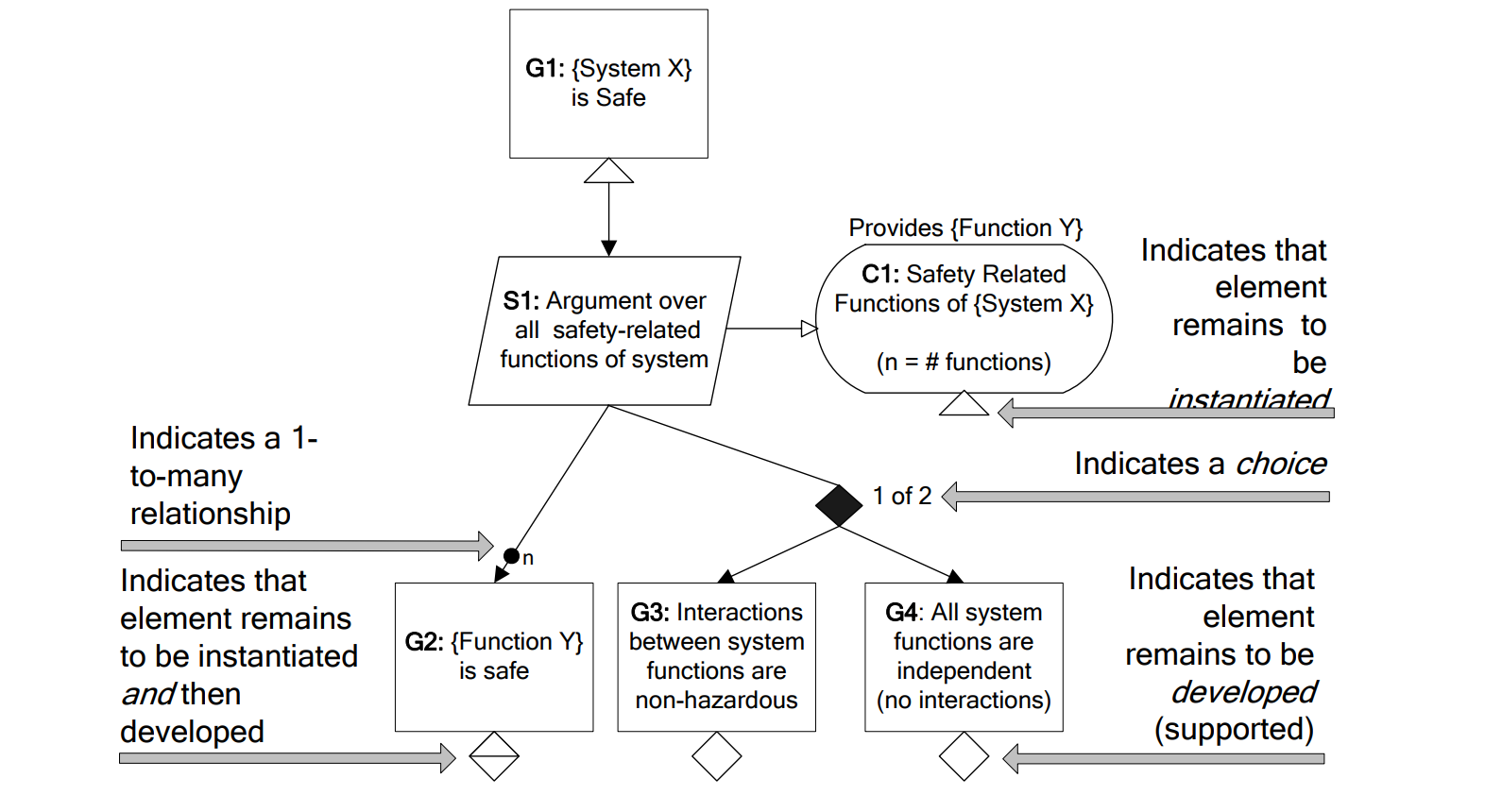}

      
      \caption{A sample assurance case pattern adapted from \cite{b1}}
      \label{fig:assurance-case-pattern}
  \end{figure}

Other GSN elements include: the \textit{Context}, the \textit{Assumption}, and the \textit{Justification} \cite{b9}. GSN elements can be decorated using the \textit{Undeveloped} decorator. This decorator is depicted as a hollow diamond applied to the bottom center of an element, indicating that a GSN element has not been developed \cite{b9}. Two main types of relationships exist between GSN elements: \textit{SupportedBy} and \textit{InContextOf} \cite{b9}. \textit{SupportedBy} is depicted as a line with a solid arrowhead and represents supporting relationships between GSN elements. \textit{InContextOf} is depicted as a line with a hollow arrowhead and represents a contextual relationship between GSN elements.  GSN has been extended to support the representation of assurance case patterns \cite{b9}. Assurance case patterns are reusable templates that facilitate the creation of assurance cases from recurring structures \cite{b9}. Figure  \ref{fig:assurance-case-pattern} describes a sample assurance case pattern represented using GSN.

\subsection{Large Language Models}
Large Language Models (LLMs) are advanced AI systems with complex neural networks, trained on vast datasets to generate human-like language and answer questions \cite{b10}. Several prompting techniques 
enable users to provide instructions and guidelines to an LLM, thereby ensuring a desired response while enhancing the efficiency and quality of its outputs \cite{b14}, \cite{b16}. For instance, the \textit{Zero-shot} technique involves prompting an LLM without providing any examples, aiming to leverage the reasoning patterns it has learned \cite{b31}. 
The \textit{CoT (Chain-of-Thought)} technique utilizes a series of intermediate reasoning steps to improve the ability of LLMs to perform complex reasoning tasks \cite{b14}.

\subsection{Using LLMs to support system assurance}
The adoption of LLMs to support system assurance tasks is increasing. 
For instance, Viger et al. \cite{b33}, Shahandashti et al. \cite{b24} as well as Gohar et al. \cite{b35} focused on the identification of defeaters using LLMs. Defeaters can be defined as \textit{"any knowledge gap that prevents complete or perfect confidence in the assurance case"} \cite{b36}. Sivakumar et al. \cite{b6} used a popular LLM i.e. GPT-4 to automatically generate safety cases. In our previous work \cite{b22}, we relied on LLMs to automatically generate assurance cases from assurance case patterns formalized using predicate-based rules.  

\section{Description of \textit{SmartGSN}}
The increasing adoption of LLMs in the system assurance field  motivated us to rely on LLMs to propose a novel tool called \textit{SmartGSN}. \textit{SmartGSN} leverages Generative AI through LLMs to semi-automate the management of assurance cases complying with GSN. \textit{SmartGSN} aims at expediting the initial drafts of assurance cases, enabling argument developers to focus on manually and interactively refining and enhancing the quality of these assurance cases. 

\subsection{Core Features of  \textit{SmartGSN}}

\subsubsection{Functional Requirements}
\textit{SmartGSN} is interactive. Its core features are:
\begin{enumerate}
    \item \textbf{Instantiation of assurance cases}: We introduced in previous work (i.e. \cite{b22}) the pattern instantiation approach \textit{SmartGSN} supports. In \cite{b22}, we used different prompting techniques to craft LLM prompts allowing us to instantiate assurance cases from assurance case patterns. Thus, our tool offers users the choice between two LLMs (i.e. GPT-4o and GPT-4 Turbo) to provide prompts that contain domain knowledge of a specific system and a given pattern in a formalized format \cite{b22}. The tool then uses that information to prompt the selected LLM to generate a system assurance case in compliance with the specified pattern. The user can then manually refine the so-generated assurance case to fix potential modeling inconsistencies.

    \item \textbf{Detection of assurance case patterns within assurance cases}: This feature aids in understanding the underlying structure of assurance cases, ensuring consistency and compliance with established patterns. Detecting these patterns also facilitates the analysis, validation, and refactoring of assurance cases, enhancing their reliability. Several researchers have used rule-based approaches and/or Machine Learning techniques to support design pattern detection  \cite{b38}, \cite{b39}. Assurance case patterns capture reoccurring modeling structures and therefore serve similar purposes as design patterns, even though they respectively target different system artifacts. In light of that analogy and inspired by existing work on design pattern detection, we rely on a metric-based rule and on generative AI (through LLMs) to detect assurance case patterns within
assurance cases. Thus, to detect a given pattern within an assurance case, each LLM (that \textit{SmartGSN} supports)   utilizes a prompt that leverages a metric-based rule. The latter aggregates \textit{n} metrics that respectively compare the similarity between an assurance case pattern and an assurance case. The textbox below illustrates that rule.   Based on that rule, if each of the \textit{n} metrics value tops its threshold, the LLM at hand concludes the pattern is detected; otherwise, the LLM concludes that the pattern is not detected. Note that, the user can set the threshold of each metric. Threshold values can also be determined empirically as in this paper.

    \item \textbf{Conversion of assurance cases from textual format to graphical format}: Assurance cases are sometimes represented in textual forms, such as structured prose, to foster freedom of expression during their creation and to make them more accessible for those who may find text easier to use and understand \cite{b8}. 
To account for the widespread use of LLMs for content and text generation, our tool facilitates the automatic conversion of LLM-generated assurance cases from textual format (structured prose) to  GSN diagrams using the GSN standard \cite {b9} guidelines.
    \item \textbf{Creation (edition) of assurance cases}: Similar to most assurance case editors, our tool features a graphical interface to facilitate the creation of assurance cases complying with GSN. The editor module is equipped with various elements, relationships, and decorators in GSN, providing drag-and-drop options on the interface for easy assurance case creation. Also, our tool supports saving work, loading ongoing projects, and exporting created assurance cases in various formats (e.g., .PNG, .JPEG, SVG, JSON). Noteworthy, as a user progressively creates an assurance case, \textit{SmartGSN} automatically and progressively generates its equivalent in structured prose. 
\end{enumerate}

\noindent\fbox{%
    \parbox{\linewidth}{%

If the value of $metric\_1$ is superior or equal to $threshold\_metric\_1$, \textbf{AND} if the value of  $metric\_2$ is superior or equal to $threshold\_metric\_2$, …,  \textbf{AND} if the value of  $metric\_n$  is superior or equal to $threshold\_metric\_n$, then conclude that the formalized assurance case pattern has been detected in the formalized assurance case. Otherwise, conclude that the formalized assurance case pattern has not been detected in the formalized assurance case.

}}

\subsubsection{Non-Functional Requirements}
Our tool supports several non-functional requirements including the following:
\begin{enumerate}
    \item \textbf{Security}: To support security, we implemented authorization with FireBase Authentication which allows secure user authentication, password management, and account creation. Credentials are managed securely on Firebase's backend, protecting sensitive data like passwords.
    \item \textbf{Accessibility}: To support accessibility in our tool, we have deployed it online, allowing users to access and utilize its features from any location with an internet connection. Responsive web design is implemented as well, allowing users from different devices to use the tool.
\item \textbf{Cross-platform compatibility}: Our tool is available online, so it can run on various operating systems, hardware platforms, and browsers.
\end{enumerate}

\subsection{Architecture of \textit{SmartGSN}}
\begin{figure}[h]
    \centering
    \includegraphics[width=0.65 \textwidth]{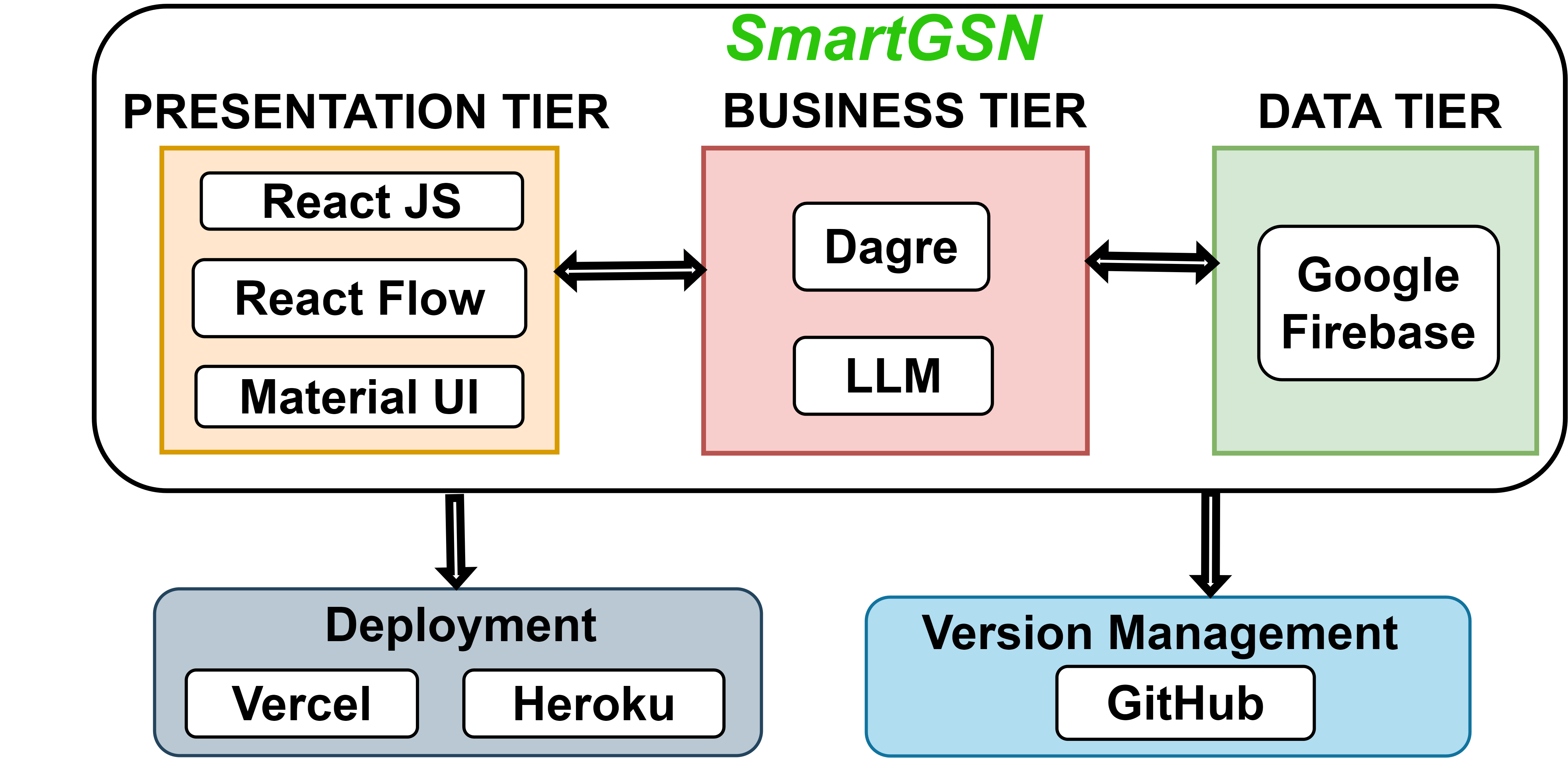}
    \caption{\textit{SmartGSN} core supporting technologies}
    \label{fig:tool_structure}
\end{figure}

To support the aforementioned features within \textit{SmartGSN}, we applied the \textit{3-tier client/server} architectural pattern to obtain the architecture of \textit{SmartGSN}. Hence, the architecture of this tool comprises three tiers: the presentation, business, and data tiers. Figure \ref{fig:tool_structure} depicts the core technologies we used to implement/deploy \textit{SmartGSN} tiers. We describe them below.

\subsubsection{Technologies used to develop the presentation tier}
The presentation tier is responsible for managing the application's user interface and experience. It allows displaying information to users and collecting their input. We used the following main technologies to implement that tier:
\begin{enumerate}
    \item \textbf{React JS}: The most popular JavaScript library for the front end in webpages. It is the core of the tool.
   We implemented React Hooks to manage performance and handle functions properly.
    
    \item \textbf{React Flow}: It is a highly customizable React library that we used for visualizing assurance cases in the graphical format i.e. GSN goal structures. 

    \item \textbf{Material UI }: It is a popular React component library that implements Google's Material Design guidelines. It provides pre-built, customizable components to help build modern, responsive, and consistent user interfaces.
\end{enumerate}

\subsubsection{Technologies used to develop the business tier} 
The business tier manages the core functionality and logic of our tool. It includes various APIs, rules, and functions designed to process user inputs and perform specific tasks, such as pattern detection and the automatic instantiation of assurance cases. 
The following technologies allowed us to implement that tier:
\begin{enumerate}
    \item  \textbf{Dagre}: A library that lays out the nodes depending on their connections. It uses various algorithms to identify the level of each node and then portrays the elements as a tree structure.
    \item \textbf{LLM (GPT and its parameters and values)}: The OpenAI API provides developers with access to powerful language models, such as GPT-4, for tasks involving natural language processing \cite{b27}. We relied on that API to create a connection between the tool and the two LLMs at hand, namely: GPT-4 Turbo and GPT-4o. After receiving a user's request containing the desired parameters and values, each LLM provides a suitable response (e.g., generate a GSN-compliant assurance case, detect a pattern in an assurance case).
\end{enumerate}

\subsubsection{Technologies used to develop the data tier}
The data tier is responsible for data storage within our tool (e.g., user credentials).  
To implement that tier, we used \textbf{Google Firebase}. The latter provides a set of tools that manage the data accessed in the back-end of the tool. Thus, \textit{SmartGSN} webpages use \textit{Authentication}, and \textit{Firestore Database} to manage passwords and registered users. 

\subsubsection{Technologies used to manage the versions of the tool}
To manage the versions of our tool, we relied on a version control platform called \textbf{GitHub}. The latter is a developing platform used for storing the various versions of projects. 

\subsubsection{Technologies used to deploy the tool}
We used two technologies to deploy \textit{SmartGSN} on the cloud:
\begin{enumerate}
    \item \textbf{Vercel}: For the deployment of the \textit{SmartGSN} front-end, we used Vercel, which is a platform designed for the deployment and hosting of web applications. Vercel automatically deploys the project to its cloud infrastructure once the build process is finished. We integrated Vercel with our GitHub repository to make updates easier when implementing the tool.
    \item \textbf{Heroku}: We deployed \textit{SmartGSN} back-end on Heroku, a cloud platform that allows us to build, run, and scale applications. Thus, we used Heroku to deploy the code that interacts with the OpenAI API. 
\end{enumerate}

\subsection{Illustration of the Tool}

\begin{figure} [h]
    \centering
    \includegraphics[width=0.95\linewidth]{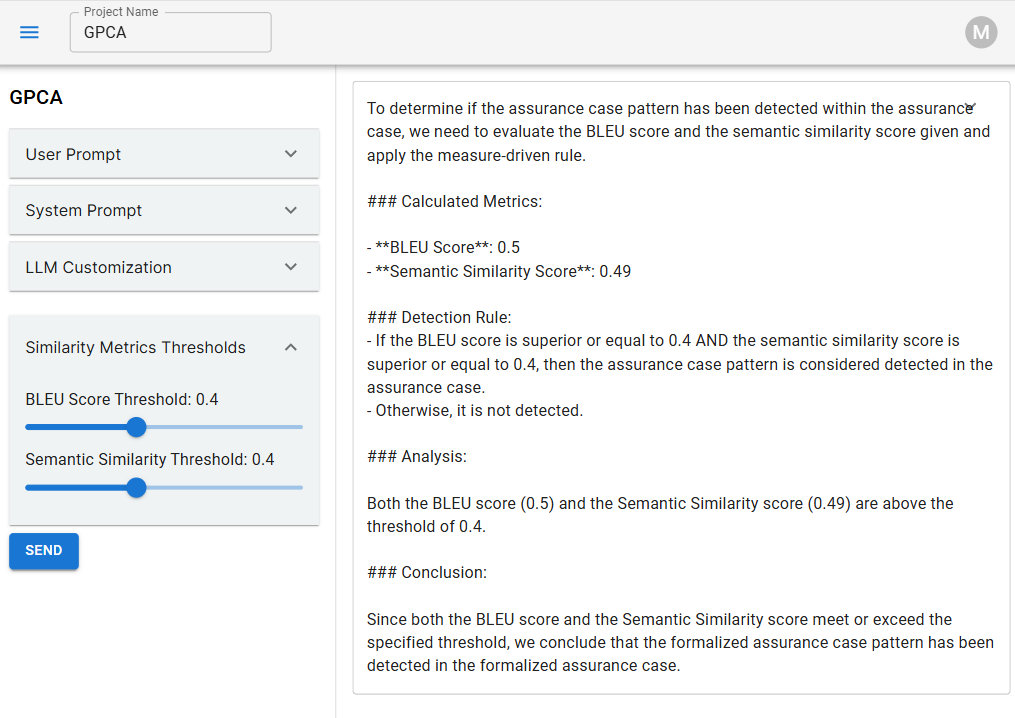}
    \caption{Partial Assurance Case Pattern Detection Interface}
    \label{fig:pattern-detection}
\end{figure}

    For brevity's sake, we only illustrate the pattern detection feature our tool supports. Figure \ref{fig:pattern-detection} provides a partial illustration of \textit{SmartGSN} pattern detection window. Thus, it depicts the collapsed interface that allows users to input values required to support pattern detection.  The LLM selected through that interface uses the supplied input values to detect a pattern in an assurance case. To detect an assurance case pattern within an assurance case, the user therefore needs to supply the values of the following input parameters: the name of the project, the LLM settings (i.e. user prompt, system prompt, and LLM customization), and metric thresholds.
Each metric threshold ranges from 0 to 1. 

\section{Empirical Evaluation}
In previous work \cite{b22}, we evaluated the LLM-based pattern instantiation technique we used to implement \textit{SmartGSN} assurance case instantiation feature. To facilitate that evaluation, we relied on \textit{SmartGSN} to automatically convert the LLM-generated assurance cases our experiments yielded from structured prose to GSN diagrams. This reduced the manual conversion effort traditionally required and enhanced the outputs' usability and practical benefits. In this paper, our evaluation focuses on the \textit{SmartGSN} pattern detection feature.
\subsection{Experiment settings}
\subsubsection{Research questions}
Our evaluation aims at answering the following research questions (RQs): 

\textbf{ (RQ1): Can \textit{SmartGSN} correctly detect assurance case patterns in assurance cases?} 

\textbf{ (RQ2): How does the choice of metric thresholds impact the  ability of \textit{SmartGSN} to detect assurance case patterns?}


\subsubsection{Dataset description}
The dataset utilized in our experiment is a collection of assurance case patterns and assurance cases manually derived (created) from these patterns. Specifically, our dataset comprises six assurance case patterns and five assurance cases related to the following five systems:

\begin{itemize}

    \item \textbf{ACAS XU}: ACAS XU (Airborne Collision Avoidance System Xu) is designed to improve the safety of drone operations by preventing collisions between drones or with other objects in their environment \cite{b41}. To ensure ACAS Xu is secure against potential threats, Zeroual et al. \cite{b41} presented a threat identification assurance case pattern. They applied this pattern to develop a partial security case specific to the ACAS Xu system.

    \item \textbf{BLUEROV2}: it is an underwater Remotely Operated Vehicle (ROV) used for inspections, underwater research, and autonomous tracking of seafloor pipelines \cite{b42}. 
    Hartsell et al. \cite{b42} created an assurance case for BlueROV2 by relying on a single pattern resulting from the combination of the ALARP (As Low As Reasonably Practicable) pattern with the ReSonAte pattern. 
    
    \item \textbf{GPCA}: The Generic Patient-Controlled Analgesia (GPCA) system, commonly known as infusion pump, is a common safety-critical system in the medical field. Lin et al. \cite{b25} proposed a safety case pattern and constructed a corresponding safety case for the GPCA system complying with this pattern. 

    \item \textbf{IM Software}: Instant messaging (IM) software facilitates information exchange, providing the basis for user interaction \cite{b43}. To ensure that the IM software is sufficiently secure, Xu et al. \cite{b43} introduced a software security top-level argument pattern to demonstrate the protection of critical assets such as user accounts and authentication information. This pattern was subsequently used in \cite{b43} to develop a security case for the IM software.
    
   \item \textbf{DEEPMIND}: The DeepMind system is the only safety-critical system in our dataset that incorporates machine learning-based components and functionalities. It uses two neural networks to predict retinal diseases from eye scans. 
   Ward and Habli. \cite{b44} introduced an assurance case pattern to justify the interpretability of machine learning in safety-critical systems. They used that pattern to create an assurance case focusing on the interpretability of DeepMind's machine learning component.
   
\end{itemize}

Our dataset consists of ACPs and ACs that are publicly available. In previous work \cite{b22}, we described how we obtained that dataset.  
Note that in our dataset, all assurance cases, except for one, are derived from a single assurance case pattern. It is only the assurance case of \textit{BlueROV2}, that is derived from a combination of two patterns.  Table \ref{table:dataset} presents statistics on our dataset, including the number of placeholders in the patterns, as well as the number of GSN elements and relationships within the patterns and assurance cases.

 \begin{table} [h]
        \centering
        \caption {Overview of our Dataset}
        \label{table:dataset}
         \begin{tabular}{|p{2.3cm} |p{2.1cm} |p{2.4cm}  |p{1.7cm}  |p{1.7cm}  |p{2.3cm}|}
        \hline
        \multirow{2}{*}{System} & \multirow{2}{*}{Domain}& \multicolumn{2}{c|}{Patterns (ACPs)} & \multicolumn{2}{c|}{Assurance Cases (ACs)}\\
        \cline{3-6}
         & & \textbf{Placeholder}& \textbf{Element}& \textbf{Element}& \textbf{Relations}\\
        \hline
        ACAS XU & Aviation  & 10 & 22 & 24 & 23 \\
        \hline
        BLUEROV2 & Automotive  & 8 & 18 & 24 & 21 \\
        \hline
        GPCA & Medical  & 21 & 23 & 27 & 26 \\
        \hline
        IM SOFTWARE & Computing  & 9 & 15 & 24 & 23 \\
        \hline
        DEEPMIND & Medical  & 26 & 17 & 23 & 23 \\
        \hline
        \end{tabular}
        \end{table}

\subsubsection{Settings (customization) of the LLMs}
Our experiment focuses on two well-known LLMs: \textit{GPT-4o} and \textit{GPT-4 Turbo}. 
To run experiments, we set each of these LLMs to its default parameters. Thus, for each LLM, we set the \textit{Temperature} to the default value of 1. Additionally, we set the default \textit{maximum number of each LLM output tokens} to be generated at 4096.

\subsubsection{Metrics used to craft pattern detection prompts}
In the context of this experiment, our metric-based rule focuses on a set of two well-known metrics (i.e. \textit{n}=2) that allow computing text similarity. These metrics are:  
\begin{itemize}

    \item \textbf{BLEU Score}: it is a widely used metric in Natural Language Processing and is one of the most common evaluation metrics for natural language texts. \textit{SmartGSN} uses that metric to assess the lexical similarity (word-by-word similarity) between an assurance case pattern and an assurance case. The BLEU score varies between 0 and 1, where 0 indicates no match between the pattern(s) and the assurance case, and 1 represents a perfect match.
    
    \item \textbf{Cosine Similarity}: \textit{SmartGSN} uses this metric to assess the semantic similarity between texts in the GSN elements of an assurance case pattern and those in an assurance case. The cosine similarity score ranges from -1 to 1, where -1 indicates completely dissimilar texts and 1 signifies identical texts.
    
\end{itemize}

To support pattern detection,  \textit{SmartGSN} uses these two metrics to compute the similarity between the assurance case pattern and the assurance case, both formalized in an advanced structured prose format we introduced in previous work \cite{b22}. \textit{SmartGSN} computes these two metrics by relying on the following Python libraries: \textit{Sacrebleu}, and \textit{scikit-learn}. Each \textit{SmartGSN} LLM then uses the values of both metrics to apply our metric-based rule when detecting patterns. Each metric threshold ranges from 0 to 1. For brevity's sake, we will only report the experiment results we obtained with the following metric thresholds: \textbf{0.2}, \textbf{0.4}, \textbf{0.6},  \textbf{0.8}, and \textbf{1}. In our experiment, we consider $threshold\_metric\_1$ and $threshold\_metric\_2$ are identical and are respectively associated with the BLEU score and the cosine similarity.

\subsubsection{Prompt description}
We rely on a combination of two well-known prompting strategies to perform our experiment \cite{b14}, \cite{b16}: \textbf{Zero-shot + CoT (Chain-of-Thought)}. When combining both techniques, we do not therefore include any examples in each LLM prompt. Still, we provide a series of intermediate logical steps needed to improve our tool's ability to detect a given pattern within an assurance case. Figures 4 and 5 illustrate the structure of our system prompt and user prompt, respectively, and can be accessed online \footnote{\url{https://github.com/Oluwafemi17/SmartGSN/blob/109ab8ed180580435442a31ee458b25aafe3cced/System_Prompt.png}} \footnote{\url{https://github.com/Oluwafemi17/SmartGSN/blob/109ab8ed180580435442a31ee458b25aafe3cced/User_Prompt.png}}.

Our LLM system prompt consists of the following elements: a sequence of intermediate steps to identify whether an assurance case pattern is present within an assurance case; context information outlining the various elements and decorators in an assurance case and an assurance case pattern represented in GSN; predicate-based rules for translating both patterns and assurance cases into a formalized format, enhancing the LLMs’ ability to understand the relationships among GSN elements in our dataset; and finally, domain information detailing the specialized knowledge and facts relevant to the system for which an assurance case pattern is to be detected. In previous work \cite{b22}, we explained how we specified context information, derived predicate-based rules and collected domain information. Our LLM user prompt includes both the formalized assurance case pattern and the formalized assurance case from which the pattern needs to be detected. Consequently, each \textit{SmartGSN} LLM relies on both system and user prompts to accurately detect a given pattern within a specific assurance case.

\subsubsection{Experiment description}
To assess the pattern detection feature of \textit{SmartGSN}, we used \textit{SmartGSN} to perform an experiment in which we ran each LLM five times 
using the aforementioned prompt. We performed that experiment five times to account for the non-deterministic nature of the LLMs \textit{SmartGSN} uses.



\subsubsection{Metrics used to assess experiment results}
To assess the pattern detection results, we relied on well-known metrics \cite{b45}: \textbf{Precision}, \textbf{Recall}, and \textbf{F-Measure}. Thus, for a given assurance case, we  compute the Precision as the number of patterns correctly detected by \textit{SmartGSN} over the total number of patterns detected by \textit{SmartGSN}. For a given assurance case, we compute the Recall as the number of patterns correctly detected by \textit{SmartGSN} over the total number of patterns manually used to create that assurance case. Finally, we compute the F-Measure as the harmonic mean of the precision and the recall. Thus, we compute F-Measure as : \textit{(2 × Precision × Recall) / (Precision+Recall)}.

\subsection{Results analysis: using \textit{SmartGSN} for pattern detection}

\subsubsection{(RQ1) \textit{SmartGSN} ability to detect patterns}

Table \ref{tab:Evaluation_result} reports the average of recall (\textbf{R}), precision (\textbf{P}), and F-Measure (\textbf{FM}) for the five systems in our dataset using two models across various thresholds. We compute each average over five runs. The "\textit{Model}" header either denotes GPT-4o or GPT-4 Turbo. 

\begin{table}
\centering
\caption{Recall (R), Precision (P), and F-Measure (FM) results}
\begin{tabular}{|l |l |l  |l |l |l |l |l |l |l |l |l |l |l |l |l |l|} \hline    
\multirow{3}{*}{System} & \multirow{3}{*}{Model} & \multicolumn{15}{|c|}{Metric Threshold} \\ \cline{3-17}
 &  & \multicolumn{3}{|c|}{0.2} & \multicolumn{3}{|c|}{0.4} & \multicolumn{3}{|c|}{0.6} & \multicolumn{3}{|c|}{0.8} & \multicolumn{3}{|c|}{1} \\ \cline{3-17} 
 &  & R & P & FM & R & P & FM & R & P & FM & R & P & FM & R & P & FM \\ \hline 
\multirow{2}{*}{ACAS XU} & GPT-4o & 1 & 1 & 1 & 1 & 1 & 1 & 1 & 1 & 1 & 0 & 0 & 0 & 0 & 0 & 0 \\ \cline{2-17}
 & GPT-4 Turbo & 1 & 1 & 1 & 1 & 1 & 1 & 1 & 1 & 1 & 0 & 0 & 0 & 0 & 0 & 0 \\ \hline 
\multirow{2}{*}{BLUEROV2} & GPT-4o & 0.5 & 1 & 0.67 & 0.5 & 1 & 0.67 & 0.5 & 1 & 0.67 & 0 & 0 & 0 & 0 & 0 & 0 \\ \cline{2-17}
 & GPT-4 Turbo & 0.5 & 1 & 0.67 & 0.5 & 1 & 0.67 & 0.5 & 1 & 0.67 & 0 & 0 & 0 & 0 & 0 & 0 \\ \hline 
\multirow{2}{*}{GPCA} & GPT-4o & 1 & 1 & 1 & 1 & 1 & 1 & 0 & 0 & 0 & 0 & 0 & 0 & 0 & 0 & 0 \\ \cline{2-17}
 & GPT-4 Turbo & 1 & 1 & 1 & 1 & 1 & 1 & 0 & 0 & 0 & 0 & 0 & 0 & 0 & 0 & 0 \\ \hline 
\multirow{2}{*}{IM Software} & GPT-4o & 1 & 1 & 1 & 0 & 0 & 0 & 0 & 0 & 0 & 0 & 0 & 0 & 0 & 0 & 0 \\ \cline{2-17}
 & GPT-4 Turbo & 1 & 1 & 1 & 0 & 0 & 0 & 0 & 0 & 0 & 0 & 0 & 0 & 0 & 0 & 0 \\ \hline 
\multirow{2}{*}{DeepMind} & GPT-4o & 1 & 1 & 1 & 1 & 1 & 1 & 0 & 0 & 0 & 0 & 0 & 0 & 0 & 0 & 0 \\ \cline{2-17}
 & GPT-4 Turbo & 1 & 1 & 1 & 1 & 1 & 1 & 0 & 0 & 0 & 0 & 0 & 0 & 0 & 0 & 0 \\ \hline

\end{tabular}
\label{tab:Evaluation_result}
\end{table}

As Table \ref{tab:Evaluation_result} shows, for each system under analysis, both \textit{SmartGSN} models achieve perfect metric scores (i.e. are able to detect patterns) at a threshold of \textbf{0.2}, except for \textit{BLUEROV2}, which has an average recall of 0.5 and an average FM-Measure of 0.67. 
At a threshold of \textbf{0.4},  \textit{ACAS XU}, \textit{GPCA}, and \textit{DeepMind}  maintain perfect metric scores, whereas the \textit{BLUEROV2}  still yields the same results. The \textit{IM Software} metric scores drop to zero at this threshold. At a  threshold of \textbf{0.8} or \textbf{1}, all systems and models achieve zero scores across all metrics, indicating an inability for \textit{SmartGSN} to detect patterns within an assurance case at these higher thresholds. Note that the values of the metrics are lower for \textit{BLUEROV2} because the assurance case of that system is derived from two patterns and not one pattern as the assurance cases of the other four systems under analysis. Still, our LLM prompts currently allow detecting a single pattern at the time. We plan to address this issue in future work.

\subsubsection{(RQ2) Impact of varying metrics thresholds on \textit{SmartGSN} pattern detection capabilities}
As Table \ref{tab:Evaluation_result} shows, the threshold values of both  metrics affect the \textit{SmartGSN} pattern detection ability; lower thresholds like \textbf{0.2} result in detecting more patterns, while at higher thresholds (e.g., \textbf{0.8}), performance drastically declines. The optimal threshold range for effective detection of patterns within assurance cases in our dataset likely falls within the \textbf{[0.2, 0.6[}  interval. Conducting more experiments in future work should confirm that observation.


\section{Conclusion and future work}
In this paper, we presented a novel tool called \textit{SmartGSN}. The latter leverages large language models (LLMs) to support the semi-automatic management of assurance cases. Our empirical evaluation focused on the pattern detection feature of that tool and showed that it can effectively rely on metrics and LLMs to detect assurance case patterns in assurance cases.

Future work will focus on refining our pattern detection  approach by exploring the use of more advanced rules and/or pattern-matching algorithms. Future work will also focus on extending our tool to support the refactoring of assurance cases using assurance case patterns. In addition, future work will focus on implementing a collaboration feature that will support the concurrent edition of the same assurance case when using \textit{SmartGSN}. We are also planning to carry out a user study to  gauge the
usability of our tool in various application domains.


\begin{credits}
\subsubsection{\ackname} The authors would like to thank the Mitacs GRI program for funding this work and Kimya K. Shahandashti for suggesting the use of React Flow. 

\subsubsection{\discintname}
The authors have no competing interests to declare. 
\end{credits}

%
%
%
\bibliographystyle{splncs04}
\bibliography{main.bib}
%

\end{document}